# Analysis of a Bloom Filter Algorithm via the Supermarket Model

Yousra Chabchoub, Christine Fricker and Hanene Mohamed

***Abstract*— This paper deals with the problem of identifying elephants in the Internet Traffic. The aim is to analyze a new adaptive algorithm based on a Bloom Filter. This algorithm uses a so-called *min-rule* which can be described as in the supermarket model. This model consists of joining the shortest queue among $d$ queues selected at random in a large number of $m$ queues. In case of equality, one of the shortest queues is chosen at random. An analysis of a simplified model gives an insight of the error generated by the algorithm on the estimation of the number of the elephants. The main conclusion is that, as $m$ gets large, there is a deterministic limit for the empirical distribution of the filter counters. Limit theorems are proved and the limit is identified. It depends on key parameters. The condition for the algorithm to perform well is discussed. Theoretical results are validated by experiments on a traffic trace from France Telecom and by simulations.**

## I. Introduction

To be efficient, network traffic measurement methods have to be adapted to the actual traffic characteristics. Internet links are currently carrying a huge amount of data at a very high bit rate (40 Gb/s in OC-768). To analyze on-line this traffic, scalable algorithms are required. They have to operate fast, using a limited small memory. The traffic is mainly analyzed at the *flow* level. A flow is a sequence of packets defined by the classical 5tuple composed of the source and destination addresses, the source and destination port numbers together with the protocol type. Flows statistics are very useful for traffic engineering and network management. In particular, information about large flows (also called elephants) is very interesting for many applications. Note that an elephant is a flow with at least $K$ packets, where $K$ is in practice equal to 20. Elephants are not numerous (around 5 to 20% of the number of flows), but they represent the main part (80-90 %) of the traffic volume in terms of packets. Elephants statistics can be exploited in various fields such as attacks detection or accounting. In the literature, some probabilistic algorithms have been developed to estimate on-line the number of elephants in a dense traffic. In [1], Flajolet analyzed the Adaptive Sampling algorithm proposed by Wegman. This algorithm is based on a special sampling method that provides a random set of the original flows. Some characteristics on elephants (number, size distribution) can be inferred from this sample. Some other algorithms (see [2], [3]) based on sampling are designed to provide elephants statistics, but they seem to be designed to very specific flow size distribution or require an a priori knowledge of the total number of flows to recover the loss of information caused by the sampling. Moreover all these algorithms are not able to identify elephants, that is to give their addresses. Such information is particularly useful for attacks detection.

Y. Chabchoub and C. Fricker are with INRIA Paris — Rocquencourt, Domaine de Voluceau, 78153 Le Chesnay, France. Email: First-Name.Last-Name@inria.fr

H. Mohamed is with Institut Universitaire de Technologie de Sceaux, 8 Avenue Cauchy, 92330 Sceaux, France, Email: Hanene.Mohamed@iut-sceaux.fr

For that, Estan and Varghese [4] propose an algorithm based on Bloom filters. This algorithm is quick enough and it uses a limited memory, but it is not adapted to traffic variations. It uses a fixed parameter which should be adjusted according to traffic intensity. Azzana in [5] then Chabchoub *et al.* [6] propose an improvement for this algorithm by adding a refreshment mechanism that depends on traffic variations. The principle of this latter algorithm is the following. The filter is composed of $d$ stages. Each stage contains $m$ counters and is associated to a hashing function. When a packet is received, its IP header is hashed by the $d$ independent hashing functions and the corresponding counter in each stage is incremented by one. When a counter reaches $K$ (the smallest elephants size), the corresponding flow is considered as an elephant. Due to the heavy Internet traffic, the filter needs to be sometimes refreshed, otherwise all the counters will exceed the threshold $K$, and then all the flows will be seen as elephants. The idea is to decrease all counters by one every time the proportion of non null counters reaches a given threshold $r$. In this way the refreshment frequency of the filter depends closely on the actual traffic intensity. Notice that the algorithm uses an improvement, the *min-rule*, also called *conservative update* in [4]. It consists in incrementing only the counters among $d$ having the minimum value, for an arriving packet. Indeed, because of collisions, the flow size is at most given by the smallest associated counters. So the min-rule reduces the overestimation of flow size. This algorithm has been first presented in [5]. In [6], a more complete version of the algorithm is developed. A new refreshment mechanism based on the average of counters values is added. In addition, the algorithm (under some modifications) is applied to attacks detection. These algorithms are validated using several traffic traces.

Chabchoub *et al.* present in [7] a theoretical analysis of the algorithm proposed by Azzana and described above. Their objective is to estimate the error generated by the algorithm on the estimation of number of elephants. The analytical study does not take into account the min-rule.

In this paper, we focus on the analysis of the min-rule. For this purpose, the algorithm described above has been slightly modified. We consider now just one filter and two hashing functions ($d = 2$). An arriving packet increments the smallest counter among the two associated counters. In case of equality, only one counter is incremented at random. In this way, every packet increments exactly one counter. A flow is declared as an elephant when its smallest associated counter reaches $C = K/d$. The same refreshment mechanism is maintained with a threshold $r$ of about 50%. The basic idea is that when the filter is not overloaded, in general, for each arriving packet of a given flow, one of the two counters will be incremented in an alternative way. It means that the two counters will have almost the same values and when the smallest one reaches $C$, the corresponding flow has a total size of about $K = dC$ packets ($C$ packets hitting each counter).

The advantage of this new algorithm is that each arriving packet increments exactly one counter. In this case the way to increment the counters is exactly, in a system of $m$ queues, the way a customer joins the shortest queue among $d$ queues chosen at random, ties being solved at random. This so-called *supermarket model* by Luczak and McDiarmid [8], [9], also known as *load-balancing model* with the choice, has been extensively studied in the literature because of its numerous useful applications. In computer science, the central result is stated in a pioneer paper by Azar *et al.* [10] then Miztenmacher [11], for a discrete time model when $n$ balls are thrown into $n$ urns with the choice. It is proved that, with probability tending to 1 as $n$ gets large, the maximum load of an urn is $\log n/\log\log n + O(1)$ when $d = 1$ and $\log\log n/\log d + O(1)$ if $d \geq 2$. Luczak and McDiarmid, in continuous time related models with the choice, explore the concentration of the maximum queue length (see [8], [12]). But the model had also already been studied in Vvedenskaya *et al.* [13], Graham [14] and others for mean-field limit theorems. In [13], a functional law of large numbers is stated: The vector of the tail proportions of queues converges in distribution as $m$ tends to infinity to the unique solution of a differential system. The differential equation has a unique fixed point $u^\rho(k) = \rho^{(d^k-1)/(d-1)}$ for a throughput $\rho$. It means that, when $d \geq 2$, the tail probabilities of the queue length decrease drastically. In Vvedenskaya and Suhov [15], variants of the choice policy and general service time distributions are investigated. Graham in [14] proves the convergence of the invariant measures to some Dirac measure. In other words, when $m$ is large, the stationary vector of the proportions of queues with $k$ customers is essentially deterministic and given by this limit.

The aim of this paper is to model the min-rule via the supermarket model and to evaluate the performance of the new proposed algorithm. In particular, we want to calculate the error generated by the algorithm on the estimation of number of elephants. Notice that this error is due to both false negatives (missed elephants) and false positives (mice considered as elephants). Let us focus on false positives. To be declared as an elephant, a mouse must be hashed to one among counters greater than $C$ after this operation. So the proportion of such counters is a good parameter to investigate in order to evaluate false positives.

The most part of the paper is the analysis of a simple model where the flows are mice of size one. It is relevant because most of the flows are mice so collisions between flows are mainly due to collisions between mice. It turns out that the probability that a mouse is detected as an elephant is bounded by the probability that a given counter is greater than $C$ just before a refreshment time. Thus the problem reduces to analyze the behavior of the model at the refreshment times. Moreover the transition phase is very short thus the study of the stationary behavior is pertinent.

The key idea of the study is to use the Markovian framework in order to rigorously establish limit theorems and analytical expressions in the stationary regime. The main result is that, as $m$ tends to infinity, the evolution of the model is characterized by a dynamical system which has at least one fixed point. When $d = 1$, this fixed point is unique and denoted by $\bar{w}$. In this paper, we conjecture its uniqueness when $d \geq 2$. The interpretation of $\bar{w}$ as a key quantity in a supermarket model with deterministic service times is discussed. Analytical expressions are given in [7] for $d = 1$, are more complicated to obtain here.

An objective would be to prove the convergence of the invariant measure of the Markov chain as $m$ tends to $+\infty$ to the Dirac measure $\delta_{\bar{w}}$ at the fixed point $\bar{w}$. In practice, such a result is completely crucial. If it is not true, if the sequence of invariant measures do not converge, the system oscillates with long periods of transition between different configurations (metastability phenomenon). So even if the algorithm performs well during a while, it can reach another state where it can give bad results. This question is partially addressed here. But the convergence of the invariant measures is conjectured, due to simulations of the algorithm where such a phenomenon has not been observed. For such a result, a possible technique is the existence of a Lyapounov function which both proves the convergence of the dynamical system to its unique fixed point $\bar{w}$ and the convergence of the sequence of invariant measures to $\delta_{\bar{w}}$. Such a function is exhibited in [7] for $d = 1$ and $C = 2$.

A simulation of the limit distribution $\bar{w}$ is done for a uniform general mice size distribution. Experiments have two goals. First to compare the original version presented in [6] and the version of the algorithm introduced here. Second, the time between two refreshments is plotted. This quantity is crucial for the trade-off between false positives and false negatives. The time to reach the stationary phase is discussed.

The organization of the paper is as follows: Section II presents the analytical results for the simple model defined to study the question of false positives. Section III is devoted to experiments. Section IV gives a discussion of the way to choose the parameter $r$ in order to have an algorithm which performs well.

## II. The Markovian urn and ball model

### A. Description of the model

In this section, the question of false positives is addressed: the probability for a mouse to be detected by the algorithm as an elephant.

The problem is studied in a simple framework, where flows are reduced to mice of size one. Thus the model can be described as a urn and ball model because one size flows hashed in a filter with $m$ counters can be viewed as balls thrown into $m$ urns with capacity $C$ under the supermarket rule: For each ball, a subset of $d$ urns is chosen at random and the ball is put in the least loaded urn, ties being resolved uniformly. Balls overflowing the capacity $C$ are rejected. If, after putting the ball, the number of non empty urns exceeds $rm$, then one ball is removed from every non empty urn.

The probability of a flow to be detected as an elephant is reduced to the probability that, after the ball arrival, all the $d$ chosen urns have $C$ balls. It is bounded by the probability that, just before a refreshment time, after putting the last ball in its urn, all the $d$ urns chosen for that contain $C$ balls. The bound is more convenient to study. The embedded model just before the refreshment times is studied.

### *B. A Markovian framework*

For fixed $C$, let us consider the sequence $(W_n^m)_{n\in\mathbb{N}}$, where $W_n^m$ denotes the vector of the proportions of urns with $0,\dots,C$ balls just before the $n$th refreshment time. For $m \geq 1$, $(W_n^m)_{n\in\mathbb{N}}$ is an ergodic Markov chain on the finite state space

$$\mathcal{P}_m^{(r)} = \{w \in \left(\frac{\mathbb{N}}{m}\right)^{C+1}, \sum_{i=0}^{C} w_i = 1,\ \sum_{i=1}^{C} w_i = \frac{\lceil rm \rceil}{m}\},$$

(where $\lceil rm \rceil$ denotes the smallest integer larger than $rm$). Thus it has a unique invariant measure $\pi_m$.

The problem is that this quantity is combinatorically intractable. Even the transition probability $P_m$ of the Markov chain is awfully difficult to write. Nevertheless, one could expect an asymptotic of this quantity when $m$ is large. In other words, the limit of the invariant measures $\pi_m$ when $m$ is large is investigated.

### *C. A dynamical system*

The way which is used here to obtain limit theorems is very classical (see [16] for example). In fact, the similar results for $d = 1$ can be found in Chabchoub et al [7]. The following results extend the case $d = 1$ to $d \geq 1$. Of course the motivation here is the case $d \geq 2$. The proofs must often be rewritten with new arguments and the sections which are still valid will be in general omitted.

The following result is that, as $m \to +\infty$, the Markov chain converges in distribution to a deterministic dynamical system which will be explicited.

Let

$$\mathcal{P} \stackrel{def}{=} \{w \in \mathbb{R}_+^{C+1}, \sum_{i=0}^{C} w_i = 1\}$$

$$\text{and } \mathcal{P}^{(r)} \stackrel{def}{=} \{w \in \mathbb{R}_+^{C+1}, \sum_{i=0}^{C} w_i = 1 \text{ and } \sum_{i=1}^{C} w_i = r\}$$

be the state spaces. Let the shift $s$ be defined as

$$s : w \mapsto (w_0 + w_1, w_2, \dots, w_C, 0) \text{ on } \mathcal{P}$$

and

$$\lambda : \mathcal{P}^{(r)} \to \mathbb{R}^+,\ w \mapsto \int_{r-w_1}^{r} \frac{du}{1-u^d}.$$

For the vector of proportions $v \in \mathcal{P}$, it is more convenient to deal with the vector of the tail proportions $u$ defined by $u_k = \sum_{i\geq k} v_i$. $G$ is then defined on $\mathcal{P}^{(r)}$ by

$$G(w) = v(\lambda(w)) \tag{1}$$

where

$(v(t))$ is associated to $(u(t))$ the unique solution of

$$u_k' = u_{k-1}^d - u_k^d,\ k \in \{1, \dots C\},\ u_0 = 1$$

with initial condition $u(0)$ corresponding to $v(0) = s(w)$.

*Proposition 1:* If $W_0^m$ converges in distribution to $w \in \mathcal{P}_m^{(r)}$ then $(W_n^m)_{n\in\mathbb{N}}$ converges in distribution to the dynamical system $(w_n)_{n\in\mathbb{N}}$ given by the recursion

$$w_{n+1} = G(w_n),\ n \in \mathbb{N}.$$

Notice that $G$ maps, by definition of $\lambda$, $\mathcal{P}^{(r)}$ to $\mathcal{P}^{(r)}$.

*Proof:* The result is a consequence of the convergence of the transition $P_m$ of the Markov chain $(W_n^m)_{n\in\mathbb{N}}$ as $m$ tends to $+\infty$ to $P$ given by

$$P(w, .) = \delta_{G(w)}.$$

It means that, starting from $w$ just before a refreshment time, at the next refreshment time, the vector of the proportions of urns tend to $G(w)$ when $m$ tends to $+\infty$. The uniform convergence stated by the following lemma provides the convenient way to prove Proposition 1.

*Lemma 1:* For $\varepsilon > 0$,

$$\sup_{w\in\mathcal{P}_m^{(r)}} P_m(w, \{w' \in \mathcal{P}_m^{(r)} : ||w' - G(w)|| > \varepsilon\}) \underset{m\to+\infty}{\to} 0.$$

*Proof:* The idea of the proof is that, starting from $w$ (with $\lceil rm \rceil$ non empty urns), after refreshment, the vector of the proportions is $s(w)$ defined by

$$s(w) = (w_0 + w_1, w_2, \dots, w_C, 0)$$

where the proportion of non empty urns is $r - w_1$. Then a number $\tau_1^m$ of balls are thrown in order to reach a state $w'$ with again $\lceil rm \rceil$ non empty urns. It has to be proved that $w'$ is close to $G(w)$. There are three steps:

1) It can be proved that this number $\tau_1^m$ is deterministic at first order, equivalent to $\lambda(w)m$, where

$$\lambda(w) = \int_{r-w_1}^{r} \frac{dt}{1-t^d} \tag{2}$$

when $m$ is large. More precisely,

$$\sup_{w\in\mathcal{P}_m^{(r)}} \mathbb{P}_w\left(\left|\frac{\tau_1^m}{m} - \lambda(w)\right| > \varepsilon\right) \underset{m\to\infty}{\to} 0. \tag{3}$$

To see it, starting from $w$, $\tau_1^m$ has an analytical expression as a sum of different numbers $Y_l$ of balls necessary to hit the $(l+1)$th non empty urn. Indeed,

$$\tau_1^m = \sum_{l=\lceil rm \rceil - w_1 m}^{\lceil rm \rceil - 1} Y_l, \tag{4}$$

where the $Y_l$s for $l \in \mathbb{N}$ are independent random variables with geometrical distributions on $\mathbb{N}^*$ with respective parameters

$$a_l = \prod_{j=0}^{l-1} \frac{l-j}{m-j},$$

i.e. $\mathbb{P}(Y_l = n) = (l/m)^{n-1}(1-l/m),\ n \geq 1$.

As $\mathbb{E}(Y_l) = 1/(1-a_l)$, computing the mean and comparing this sum with integrals leads to

$$\sup_{w \in \mathcal{P}_m^{(r)}} \mathbb{E}_w \left( \frac{\tau_1^m}{m} \right) \underset{m \to \infty}{\to} \lambda(w). \tag{5}$$

At the same time, as $\mathrm{Var}(Y_l) = a_l/(1-a_l)^2$,

$$\sup_{w \in \mathcal{P}_m^{(r)}} \frac{\mathrm{Var}_w(\tau_1^m)}{m} \underset{m \to \infty}{\to} \int_{r-w_1}^{r} \frac{dt}{(1-t^d)^2}. \tag{6}$$

By Bienaymé-Chebychev's inequality, using equations (5) and (6), it proves (3).

2) From the previous fact, there is a natural coupling throwing $\tau_1^m m$ balls or $\lambda(w)m$ balls where the vectors of proportions $W_1^m$ and say $\tilde{W}_1^m$ are close to each other. Then there is also a coupling throwing $\lambda(w)m$ and a Poisson random variable with parameter $\lambda(w)m$, for which, by Chernoff's inequality, the vector of proportions $\tilde{W}_1^m$ and say $\hat{W}_1^m$ are close.

3) The vector of proportions $\tilde{W}_1^m$ obtained by coupling is the vector of proportions at time $\lambda(w)$ in a queueing supermarket model without departures. The model consists of $m$ queues with capacity $C$ where customers arrive according to a Poisson process with rate $m$. At each arrival, a subset of $d$ queues is chosen and the customer joins the shortest one, ties being solved at random. Let $W^m(t)$ be the vector of the proportions of $m$ queues with $0, 1, \ldots, C$ customers at time $t$. It is more convenient to deal with the tail proportions defined as

$$U_k^m(t) = \sum_{i \geq k} W_i^m(t).$$

Given $W^m(0) = s(w)$, we have that

$$\hat{W}_1^m = W^m(\lambda(w)).$$

By the convergence of the Markov process $(U^m(t))$ to the fluid limit (see Vvedenskaya et al. [13]), it holds that $\hat{W}_1^m$ converges in distribution to $v(\lambda(w))$ where $v$ is associated to $u$ the fluid limit of $(U^m(t))$, the unique solution of the differential system

$$\frac{du_k}{dt} = u_{k-1}^d - u_k^d \ (1 \leq k \leq C),\ u_0 = 1, \tag{7}$$

with initial condition $u(0)$ corresponding to $v(0) = s(w)$. Moreover using the continuity of a solution of a differential equation with respect to the initial condition, for each $\varepsilon,\ t > 0$, $0 \leq k \leq C$,

$$\mathbb{P}( \sup_{w \in \mathcal{P}_m^{(r)}} |U_k^m(\lambda(w)) - u_k(\lambda(w))| > \varepsilon) \underset{m \to \infty}{\to} 0$$

which straightforwardly leads to

$$\mathbb{P}( \sup_{w \in \mathcal{P}_m^{(r)}} \| \hat{W}_1^m - G(w) \| > \varepsilon) \underset{m \to \infty}{\to} 0$$

where

$$G(w) = v(\lambda(w)).$$

It ends the proof of the lemma. ∎

The argument to obtain Proposition 1 from Lemma 1 is standard and detailed in [7, Proposition 1]. It is omitted here. ∎

### D. Fixed point of the dynamical system

The function

$$\begin{aligned} \mathcal{P}^{(r)} &\longrightarrow \mathcal{P}^{(r)} \\ w &\longmapsto G(w) \end{aligned}$$

being continuous on the convex compact set $\mathcal{P}^{(r)}$, by Brouwer's theorem, it has a fixed point.

It remains to prove the uniqueness of the fixed point. Recall that, for $d = 1$, the proof is based on the interpretation of the fixed point equation

$$G(w) = w$$

as the invarriant measure equation

$$wP = w$$

of some ergodic Markov chain with transition $P$. This Markov chain is the queue length at the service time completions of a $M/G/1/C$ queue with deterministic service times equal to 1 with arrival rate $\lambda(w)$. The proof of the uniqueness of the solution of $w = \mu_{\lambda(w)}$ is then based on the coupling argument that, if $\lambda \leq \lambda'$ then $\mu_\lambda$ is stochastically dominated by $\mu_{\lambda'}$ (see [7] for details).

Let us try to extend the argument for the case $d \geq 2$. For that, let us consider the following system. Balls are thrown into $m$ urns with capacity $C$ with a Poisson arrival process with rate $\lambda m$. Each ball joins the least loaded urn among a subset of $d$ urns, chosen at random. The ties are resolved uniformly. At each unit time, one ball is removed from each non-empty urn. It can be proved as in Proposition 1 that the vector of the proportions of urns with $k$ balls just before time $n$ converges, when $m$ is large, to a dynamical system

$$w_{n+1} = H(w_n)$$

where, for $v$ defined as previously with initial condition $v(0) = s(w)$,

$$H(w) = v(\lambda).$$

But the argument fails for $d \geq 2$. Indeed, the equation $w = H(w)$ can not be interpreted as the invariant measure equation of some ergodic Markov chain $(L_n)_{n \in \mathbb{N}}$ on $\{0, \ldots, C\}$ because the differential system (7) is not linear for $d > 1$. If it was then there should exist $P_\lambda$ such that

$$H(w) = v(\lambda) = v(0)P_\lambda = s(w)P_\lambda = wP$$

where $P = QP_\lambda$ because $s(w)$ can easily be written $wQ$ where $Q$ is a transition matrix.

Thus another way to prove it should be found and, at this point, the uniqueness of the fixed point is conjectured.

### E. Identification of the fixed point

If the capacity is infinite, then the parameter $\lambda(\bar{w})$ is equal to $r$. In this case, it is simple to have the explicit expression of $\bar{w}_1$, which is a good approximation for the case $C = 20$. It is the purpose of this section.

Assume that $C = +\infty$. By definition,

$$\lambda(\bar{w}) = \int_{r-\bar{w}_1}^{r} \frac{dt}{1-t^d}$$

where $F(x) = \int_0^x \frac{dt}{1-t^d}$ defines a bijection from $[0, 1[$ to its image. Using that $\lambda(\bar{w}) = r$ for $C = +\infty$, it can be rewritten

$$r = F(r) - F(r - \bar{w}_1),$$

or

$$\bar{w}_1 = r - F^{-1}(F(r) - r). \tag{8}$$

Notice that for $d \in \mathbb{N}$, $F$ has an explicit expression. For $d = 1$, $F(x) = -\log(1-x)$ which leads (see [7]) to

$$\bar{w}_1 = (1-r)(e^r - 1).$$

Moreover, for $d = 2$, $F(x) = \mathrm{argth} x$ and thus

$$\bar{w}_1 = r - \frac{r - \mathrm{th} r}{1 - r \mathrm{th} r}. \tag{9}$$

In Figure 1, $\bar{w}_1$ is plotted for $d = 1,\ 2$.

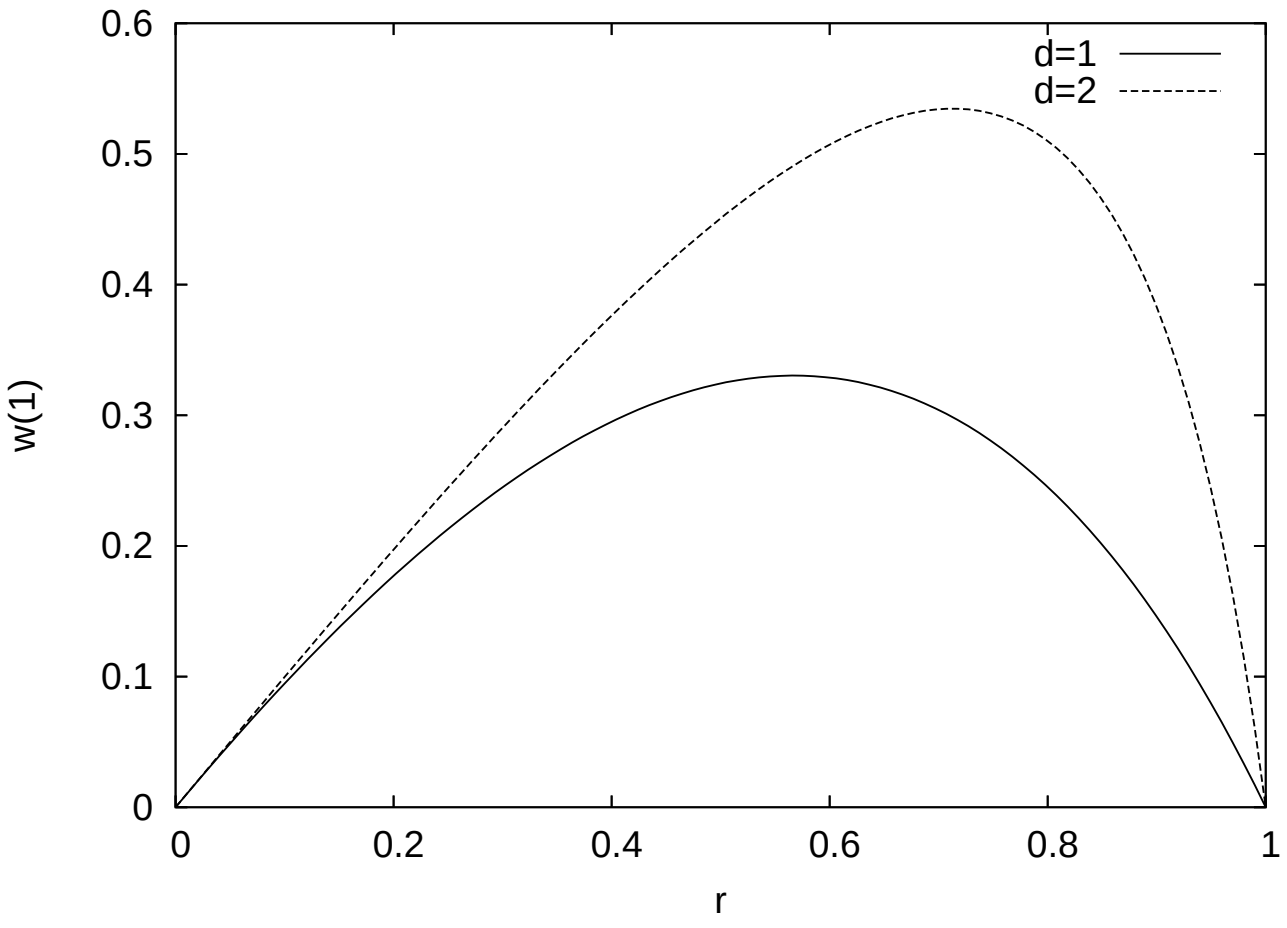

Fig. 1. Limit proportion $\bar{w}_1$ of counters with value 1 for $d = 1, 2$.

### F. Convergence of invariant measures

A first result is obtained. It is proved in [7, Proposition 2] and recalled here omitting the proof.

*Proposition 2:* Let, for $m \in \mathbb{N}$, $\pi_m$ be the stationary distribution of $(W_n^m)_{n\in\mathbb{N}}$. Define $P$ as the transition on $\mathcal{P}^{(r)}$ given by $P(w, .) = \delta_{G(w)}$. Any limiting point $\pi$ of $(\pi_m)_{m\in\mathbb{N}}$ is a probability measure on $\mathcal{P}^{(r)}$ which is invariant for $P$ i.e. that satisfies $G(\pi) = \pi$.

As noticed in the introduction, the limiting point of $\pi_m$ is not necessarily unique, because there is not a unique measure $\pi$ such that $G(\pi) = \pi$. Nevertheless $G$ has a unique fixed point thus $G(\delta_{\bar{w}}) = \delta_{\bar{w}}$. But imagine that $G$ has cycles, i.e. that there exist $n \geq 2$ and $w_1, \ldots, w_n$ in $\mathcal{P}^{(r)}$ such that

$$G(w_i) = w_{i+1}\ (1 \leq i < n),\ G(w_n) = w_1$$

then $\pi = 1/n \sum_{i=1}^n \delta_{w_i}$ is invariant under $G$. It gives two different limiting points for $\pi_m$. A way to prove the convergence is to exhibit a Lyapounov function for $G$ (see [7, Theorem 1] for details). Such a Lyapounov function is exhibited in [7] for $d = 1$ and $C = 2$. It is not investigated here.

### G. General mice size distribution

The aim of the subsection is to extend the previous results to a model with general size distribution. An approximated model is taken. Indeed, as mice size are short (with mean close to some units, in real traffic traces, close to 4), an approximated model is to consider that the packets of the mice are thrown without interleaving in the target counters. It means that the packets of the different mice arrive consecutively in the filter.

The model chosen is thus an urn and ball model where balls are thrown by batches . The balls in a batch are thrown together in a unique urn, the least loaded urn among $d$ chosen at random in the $m$ urns. The $i$th batch is composed with $S_i$ balls, where the $S_i$s are independent random variables with distribution denoted by $p$.

Let also $(W_n^m)_{n\in\mathbb{N}}$ be the sequence of vectors giving the proportions of urns at $0, \ldots, C$ just before the $n$th refreshment time in this model where balls are thrown by batches. The dynamic is the same: If, before a refreshment time, the state is $w \in \mathcal{P}_m^{(r)}$, it becomes $s(w)$ and then a number $\tau_1^m(w)$ defined by (4) of successive batches are thrown in urns until $\lceil rm \rceil$ urns are non empty. The model generalizes the previous one obtained for mice of size one ($p(1) = 1$).

Note that equation (9) is extended in this case by

$$\bar{w}_1 = r - \frac{r - \mathrm{th}(\mathrm{r}/\mathbb{E}(\mathrm{S}))}{1 - r\mathrm{th}(\mathrm{r}/\mathbb{E}(\mathrm{S}))}. \tag{10}$$

Let $G$ be defined on $\mathcal{P}$ by

$$G(w) = v(\lambda(w)) \tag{11}$$

where the tail function $(u(t))$ corresponding to $(v(t))$ is the unique solution of the differential equation

$$\frac{du_k}{dt} = \sum_{j=1}^{k} p_j (u_{k-j}^d - u_k^d)\ (1 \leq k \leq C),\ u_0 = 1. \tag{12}$$

Everything in Section II remains valid. The supermarket model obtained by coupling is a model with batch arrivals without departures. Its mean-field limit is obtained as previously and leads to the differential equation (12). Propositions 1 and 2 hold. The description of the unique fixed point can be extended.

## III. Experiments

In this section, the proposed algorithm is tested against an ADSL traffic trace from France Telecom IP backbone network. This traffic trace has been captured on a Gigabit Ethernet link in October 2003 between 9:00 pm and 10:00 pm. This period corresponding to a peak activity by ADSL customers, its duration is 1 hour and contains more than 10 millions of TCP flows.

In our experiments, the filter consists of $m = 2^{20}$ counters associated to two independent hashing functions $(d = 2)$. Elephants are here defined as flows with at least 20 packets $(K = 20)$.

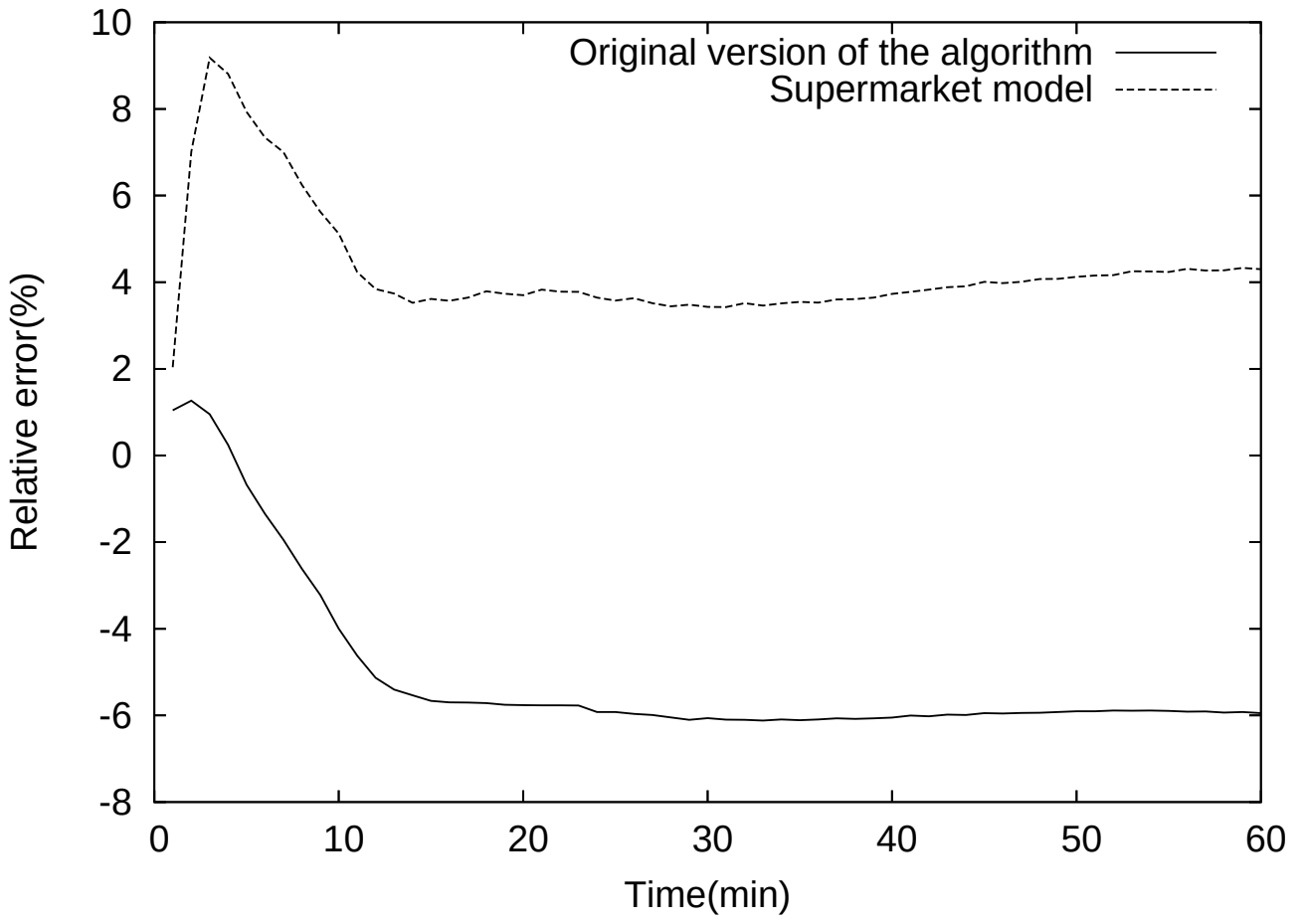

Fig. 2. Impact of the Supermarket model on the estimation of number of elephants , $r = 50\%$, France Telecom trace

The relative error on the estimated number of elephants is plotted in Figure 2. Two different versions of the algorithm are considered: The original algorithm developed in [5], [6] and the proposed algorithm using the supermarket model. We recall, that these two algorithms use the min-rule (incrementing only the smallest counter), but in a different way: In case of equality, only one counter is incremented at random with the supermarket model whereas, the two counters are incremented in the original version of the algorithm. Results show that both methods give a good estimation of number of elephants, for the whole duration of the trace.

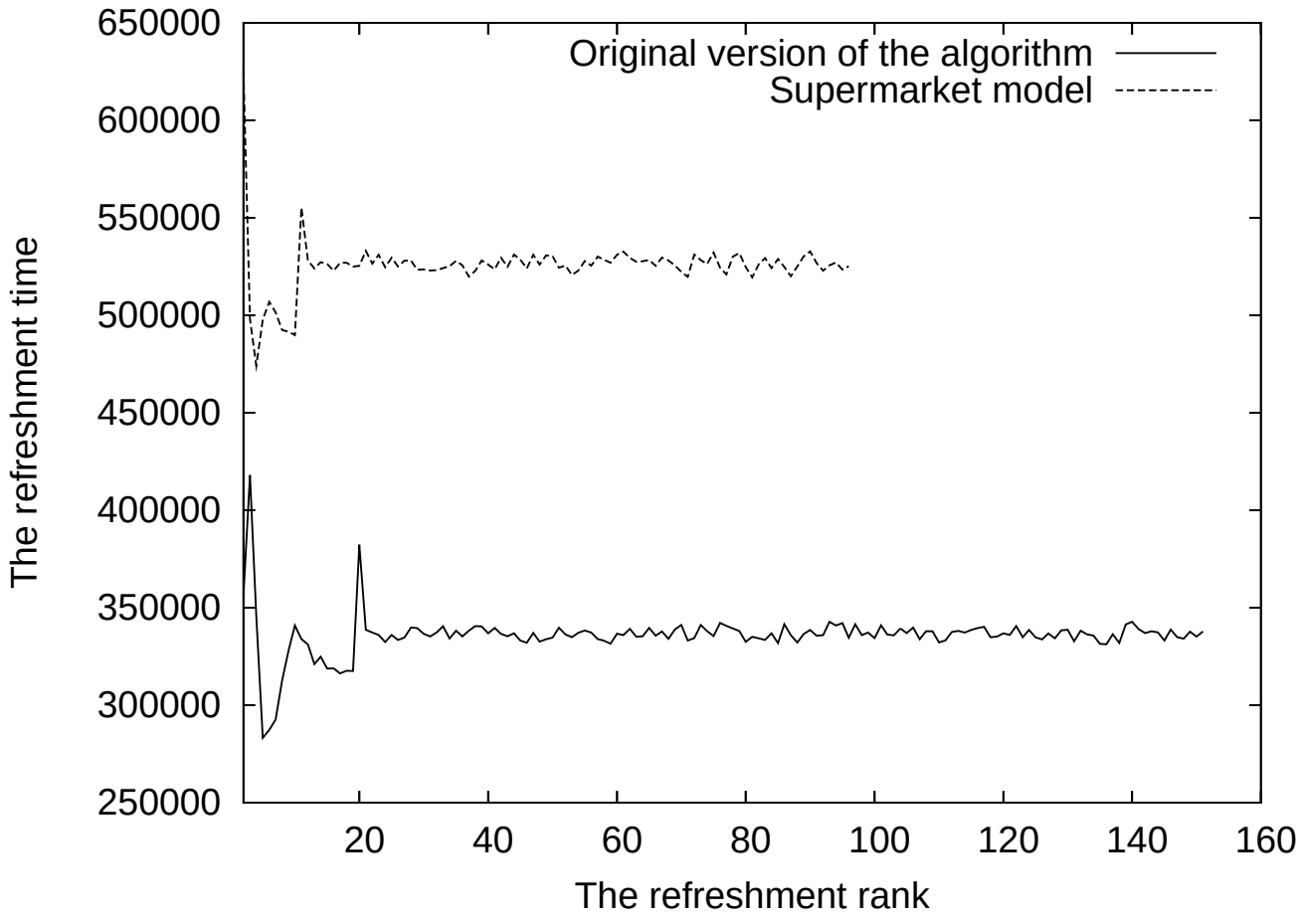

Fig. 3. Duration of the transition and the stationary phase, $r = 50\%$, France Telecom trace

Figure 3 presents *the inter-refreshment time* (duration between two successive refreshments in terms of number of arriving packets) for the whole traffic trace. It can be noticed that the stationary phase is reached at the $K$th refreshment time. So the transition phase seems to be rather short, according to experiments. The stationary inter-refreshment time using the algorithm based on the supermarket model is higher than the one obtained with the original version of the algorithm. This can be explained by the fact that with the supermarket model every arriving packet increments exactly one counter, whereas in the original version, if the two selected counters are equal, they are both incremented by one. In particular, when they are both null, they will be both impacted. As a consequence, the proportion of non null counters grows faster and the filling up threshold $r$ is reached more quickly.

Figure 3 gives an explanation to the behavior of the algorithms plotted in Figure 2. In the original algorithm, the inter-refreshment time is lower thus more elephants are missed. The error is thus negative. In the supermarket version, the error is positive due to false positives.

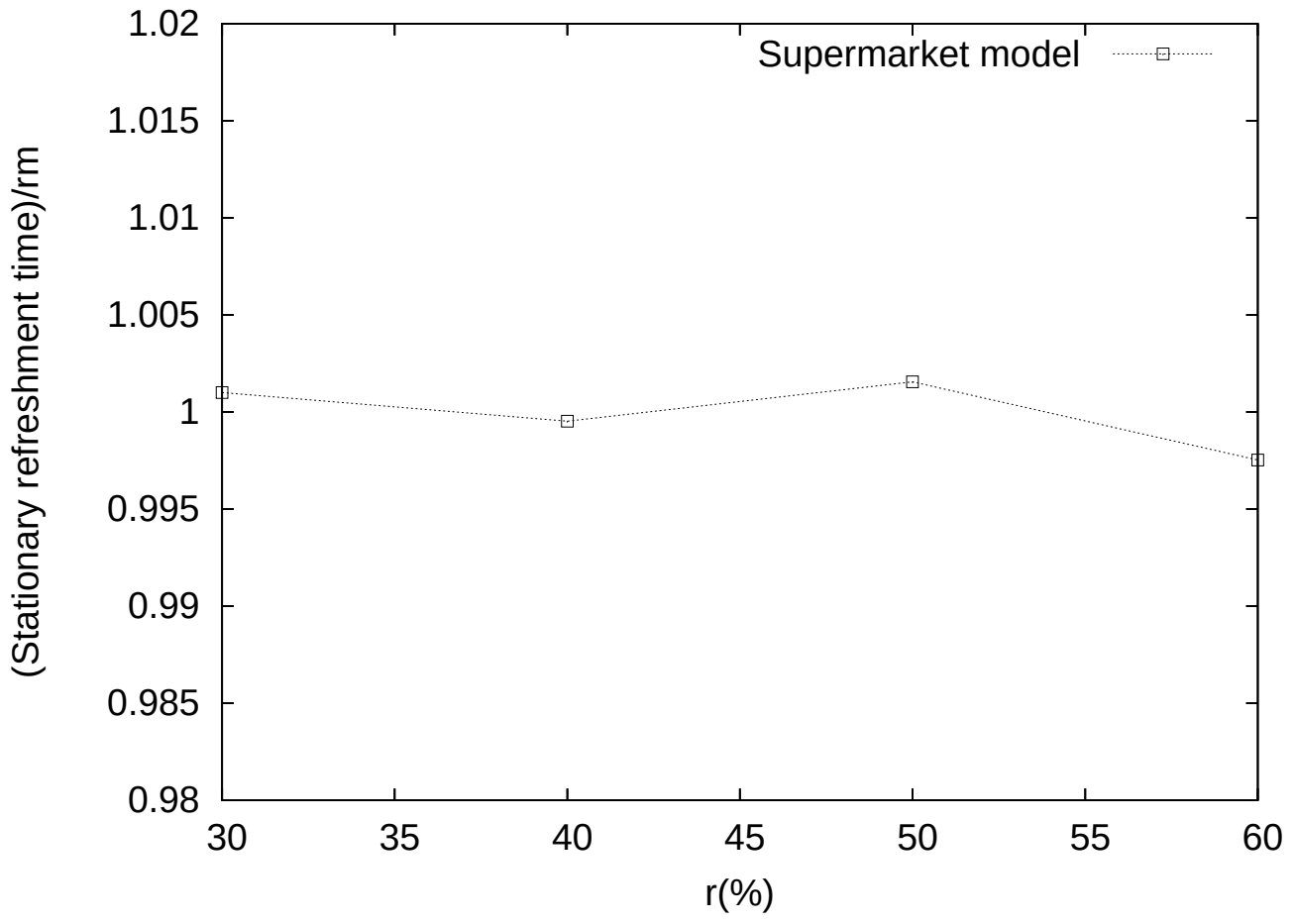

Fig. 4. Comparison between $rm$ and the stationary inter-refreshment time $\tau_\infty^m$, France Telecom trace

In Figure 4, the impact of $r$ on the stationary inter-refreshment time $\tau_\infty^m$ is investigated. More precisely $\tau_\infty^m/rm$ is plotted for various values of $r$. According to experiments, $\tau_\infty^m$ is very close to $rm$. In fact the refreshment can be seen as removing $rm$ from the sum $S$ of all counters (decreasing by one all non null counters which are exactly $rm$ as the refreshment is performed as soon as the filling up threshold $r$ is reached). As we are in the stationary phase, we have convergence of $w_i$, the proportion of counters at $i$, for $i \in \{0, \ldots, C\}$. Therefore the sum of all counters converges. So just before the next refreshment, $rm$ packets must be inserted into the filter, to let $S$ have its former value. Packets belonging to elephants which have been detected are not taken into account. Those packets are very numerous and they are not inserted into the filter to avoid polluting it.

## IV. Discussion

The performance of the algorithm clearly depends on the filling up $r$. To have a good estimation of the number of elephants , $r$ must be around $50\%$. When $r$ has higher values, elephants number will be largely overestimated due to false positives. The key quantity is $\bar{w}_i$, the stationary proportion of counters at $i$ when $m$ gets large. An explicit expression for $\bar{w}$ is not available even if a numerical value could be computed. Nevertheless, less ambitiously, one can maybe simply found the critical value of $r$ for which $\bar{w}_C$ gets non negligible. At least, the impact of $r$ on $\bar{w}$ is shown here by simulation.

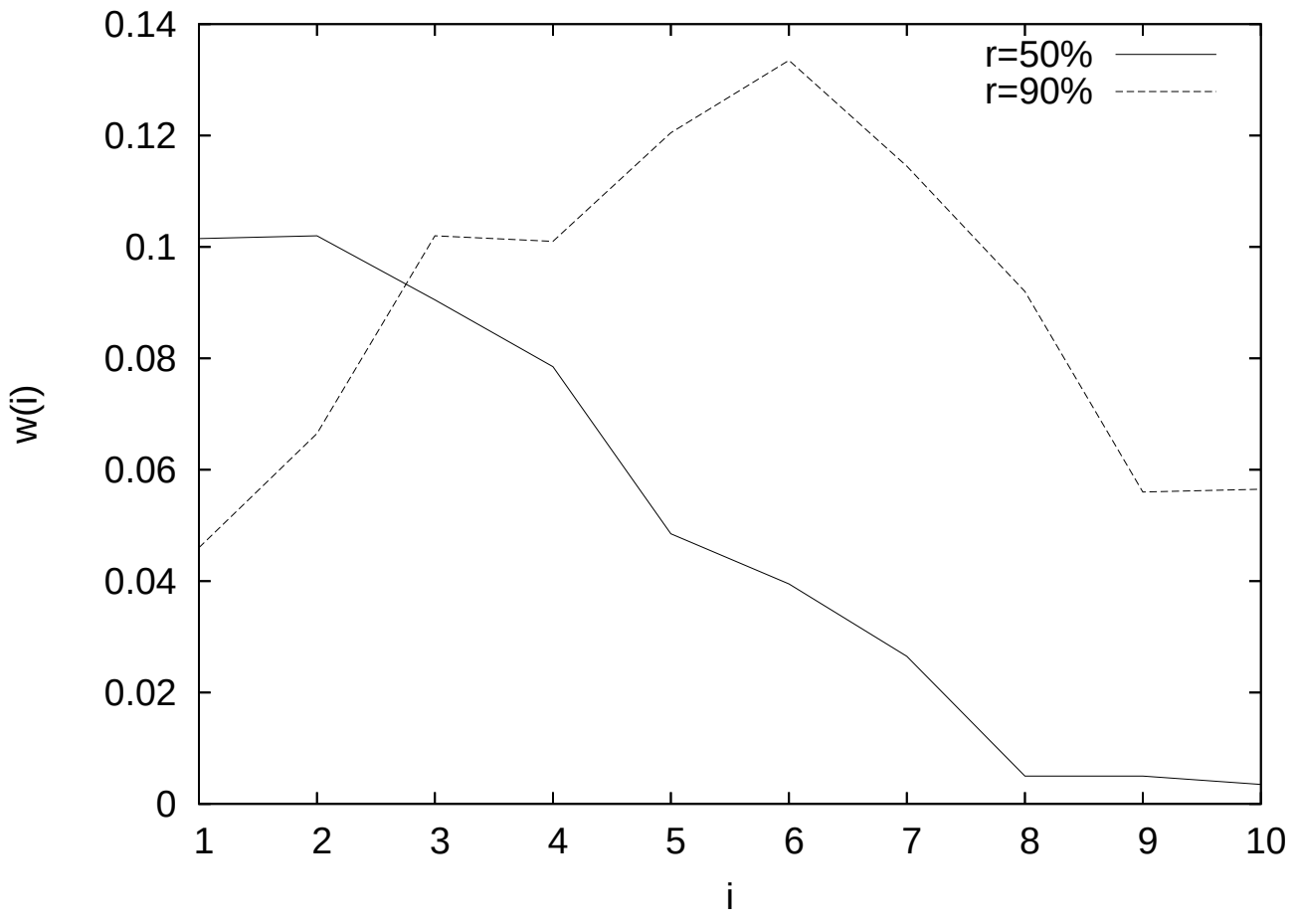

Fig. 5. Impact of $r$ on the limit stationary distribution $\bar{w}$, simulation using only mice with uniform size distribution of mean 4, $m = 2000$

Figure 5 ($w$ is written for $\bar{w}$) is not based on a real traffic trace but on simulation. The objective here is to evaluate the limit stationary distribution $\bar{w}$ if we consider a traffic composed only of mice. The mice mean size is taken equal to four to be close to the real traffic (This value is deduced from the real traffic trace). Under these conditions, we obtain a decreasing limit stationary distribution of $\bar{w}$, when $r$ equals $50\%$. For a filling up threshold of $90\%$, counters are very likely to be higher. We can notice that the main part of counters values is around six and there are many counters at $C$. This explains the fact that with a filling up threshold around $50\%$, the algorithm performs better. Equation (10) with $C = \infty$ gives a very good approximation of the values of $\bar{w}_1$ in this case. Indeed, the analytical expression gives $\bar{w}_1 = 0.10$ for $r = 0.5$ and $\bar{w}_1 = 0.05$ for $r = 0.9$. These values are very close to the values obtained on Figure 5 by simulation.

## V. Conclusion

We analyze in this paper a new algorithm catching on-line elephants in the Internet. This algorithm is based on Bloom filters with a refreshment mechanism that depends on the current traffic intensity. It also uses a conservative way to update counters, called the min-rule. This latter is exactly to increment the lowest counter among a set of $d$ chosen at random as in the supermarket model which provides a much lower tail distribution for the counter values. For a model involving just mice, limit theorems investigate the existence of a deterministic limit for the empirical distribution of counters values, when the filter size gets large. This limit can be exploited to adjust the parameters for the algorithm to perform well. The accuracy of the algorithm and some theoretical results are tested against a traffic trace from France Telecom and by simulations.

## References

[1] P. Flajolet, "On adaptative sampling," *Computing*, pp. 391–400, 1990.

[2] O. Gandouet and A. Jean-Marie, "Loglog counting for the estimation of ip traffic," in *Proceedings of the 4th Colloquium on Mathematics and Computer Science Algorithms, Trees, Combinatorics and Probabilities*, Nancy, France, 2006.

[3] Y. Chabchoub, C. Fricker, F. Guillemin, and P. Robert, "Inference of flow statistics via packet sampling in the internet," *IEEE Communication Letters*, pp. 897 – 899, 2008.

[4] C. Estan and G. Varghese, "New directions in traffic measurement and accounting," in *Proc. Sigcomm'02*, Pittsburgh, Pennsylvania, USA, August 19-23 2002.

[5] Y. Azzana, "Mesures de la topologie et du trafic internet," Ph.D. dissertation, Université de Paris 6, July 2006. [Online]. Available: http://www-c.inria.fr/twiki/pub/RAP/FormerMembers/Azzana-PhD.pdf

[6] Y. Chabchoub, C. Fricker, F. Guillemin, and P. Robert, "Adaptive algorithms for identifying large flows in ip traffic," in *Submitted to ITC21 21st International Teletraffic Congress*, september 2009.

[7] Y. Chabchoub, C. Fricker, F. Meunier, and D. Tibi, "Analysis of an algorithm catching elephants on the internet," in *Fifth Colloquium on Mathematics and Computer Science*, ser. DMTCS Proceedings Series, september 2008, pp. 299–314.

[8] M. J. Luczak and C. McDiarmid, "On the maximum queue length in the supermarket model," *Ann. Probab.*, vol. 34, no. 2, pp. 493–527, 2006.

[9] ——, "Asymptotic distributions and chaos for the supermarket model," *Electron. J. Probab.*, vol. 12, pp. no. 3, 75–99 (electronic), 2007.

[10] Y. Azar, A. Z. Broder, and A. R. Karlin, "On-line load balancing," *Theoret. Comput. Sci.*, vol. 130, no. 1, pp. 73–84, 1994.

[11] Mitzenmacher, "The power of two choices in randomized load balancing," Ph.D. dissertation, Berkeley, 1996.

[12] M. J. Luczak and C. McDiarmid, "On the power of two choices: balls and bins in continuous time," *Ann. Appl. Probab.*, vol. 15, no. 3, pp. 1733–1764, 2005.

[13] N. D. Vvedenskaya, R. L. Dobrushin, and F. I. Karpelevich, "A queueing system with a choice of the shorter of two queues—an asymptotic approach," *Problemy Peredachi Informatsii*, vol. 32, no. 1, pp. 20–34, 1996.

[14] C. Graham, "Chaoticity results for "join the shortest queue"," in *Council for African American Researchers in the Mathematical Sciences, Vol. III (Baltimore, MD, 1997/Ann Arbor, MI, 1999)*, ser. Contemp. Math. Providence, RI: Amer. Math. Soc., 2001, vol. 275, pp. 53–68.

[15] N. D. Vvedenskaya and Y. M. Sukhov, "Dobrushin's mean-field approximation for a queue with dynamic routing," INRIA, Tech. Rep. 3328, dec 1997.

[16] V. Dumas, F. Guillemin, and P. Robert, "A Markovian analysis of additive-increase multiplicative-decrease algorithms," *Adv. in Appl. Probab.*, vol. 34, no. 1, pp. 85–111, 2002.